\documentstyle[12pt]{article} 
\newcommand{\be}{\begin{equation}} 
\newcommand{\en}{\end{equation}}
\newcommand{\bea}{\begin{eqnarray}}
\newcommand{\ena}{\end{eqnarray}}

\newcommand{\hbo}{\hbox to 1 true cm {\hfill } } 
\newcommand{\tr}{\hbox{tr}}

\def\dslash{\partial\kern-.5em\slash}
\def\kslash{k\kern-.5em\slash}
\def\pslash{p\kern-.5em\slash}

\begin{document} 
\vglue 1truecm
  
\vbox{ 
\hfill February 7, 1996 
}
  
\vfil
\centerline{\large\bf The quark condensate in the GMOR relation }

\bigskip
\centerline{ K.\ Langfeld, C.\ Kettner } 
\bigskip
\vspace{1 true cm} 
\centerline{ Institut f\"ur Theoretische Physik, Universit\"at 
   T\"ubingen }
\centerline{D--72076 T\"ubingen, Germany.}
\bigskip
\vskip 1.5cm

\begin{abstract}
\noindent 
The quark condensate which enters the 
Gell-Mann-Oakes-Renner (GMOR) relation, is investigated in the 
framework of one-gluon-exchange models. The usual definition of 
the quark condensate via the trace of the quark propagator produces 
a logarithmic divergent condensate. In the product of current mass 
and condensate, this divergence is precisely compensated by the bare 
current mass. The finite value of the product in fact does not contradict 
the relation recently obtained by Cahill and Gunner. Therefore 
the GMOR relation is still satisfied. 

\end{abstract}

\vfil
\hrule width 5truecm
\vskip .2truecm
\begin{quote} 
$^*$ Supported in part by DFG. 
\end{quote}
\eject

The common believe that Quantumchromodynamics (QCD) is the right theory 
of strong interactions, stems from a good agreement of theoretical 
predictions with the experimental data in deep inelastic scattering 
experiments~\cite{yn83}. The theoretical approach is thereby based 
upon perturbation theory, which is the appropriate technique since 
the effective expansion parameter, i.e. the running coupling constant, 
is small at high momentum transfer. At medium energies, the non-trivial 
ground state properties of the QCD vacuum induce corrections to the 
perturbative result. These corrections can be systematically included 
within an operator-product-expansion (OPE)~\cite{shi79,tar84}, where 
non-perturbative vacuum properties are parameterized in terms of 
condensates. The powerful technique of QCD sum rules~\cite{nar89,rei85} 
combines the OPE and dispersion relations in order 
to address hadron properties. The actual values of the condensates 
are obtained by fitting the masses of mesons and hadrons~\cite{nar89,rei85}. 
Unfortunately, a direct calculation of the condensates and hadron 
properties from QCD is not feasible at present stage, since a large 
effective coupling constant at low energies renders the investigations 
cumbersome. 

Exploiting the realization of symmetries in low energy QCD provides 
further inside into the QCD vacuum structure and results in low 
energy theorems~\cite{ch84}. In particular, the spontaneous breakdown 
of chiral symmetries~\cite{nam61} leads to a good description of the light 
pseudo-scalar mesons as Goldstone bosons~\cite{appli}. A further 
consequence of 
chiral symmetry is the Gell-Mann-Oakes-Renner (GMOR) relation, which 
relates the pion mass $m_\pi $ and the pion decay constant $f_\pi $ 
to the product of current quark mass $m$ and quark condensate 
$\langle \bar{q}q \rangle$. In the SU(2) iso-spin symmetry case, one has 
\be 
m_{\pi } ^{2} \, f_{\pi }^{2} \; = \; - 2 m \langle \bar{q} q \rangle \; . 
\label{eq:1} 
\en 
This equation is of particular 
importance, since it provides access to meson observables by 
simply studying properties of the quark ground state. Recently, 
the GMOR relation gained further phenomenological importance in 
estimating the density dependence of the meson observables as
requested to describe compact star matter~\cite{newbr}. 

Despite of the success due to sum rules and symmetry arguments, one must 
resort to low energy effective models in order to move towards a 
``microscopic'' descriptions of baryons and mesons. The one-gluon-exchange 
models~\cite{pa77,sme91,rob94,fra96,cah92,cah95} have proven to be 
phenomenological very successful. In these models, the quark interaction 
is described by an effective one-gluon exchange, which match the well 
known interaction at high energies. The models are (at least one-loop) 
renormalizable, and confinement of quarks can be described by a quark 
propagator which finally vanishes due to infra-red 
singularities~\cite{sme91}, or which does not have poles corresponding to 
asymptotic quark states~\cite{rob94}. Meson phenomenology can be 
successfully addressed~\cite{fra96,cah95} (for a review see~\cite{rob94}), 
and the description of baryons 
in terms of the Faddeev equation approach seems feasible~\cite{cah92}. 

Within the context of the one-gluon-exchange model, Cahill and Gunner 
recently argued that the GMOR relation (\ref{eq:1}) is wrong. A new 
formula which determines the product of pion mass and decay constant in 
terms of the quark propagator, was proposed~\cite{cah96,fra96}. The 
discrepancy originates from the definition of the quark condensate as 
the trace of the 
quark propagator. This definition naturally provides a logarithmic 
divergent condensate, whereas the GMOR relation seems to require 
a finite condensate. The arguments involved are not confined to the 
one-gluon-exchange model implying that the problem at hand exists in 
the general context of QCD. 

\parindent=12pt
In this letter, we reconsider the GMOR relation in the context of 
the one-gluon-exchange models. We will show that, in the product of 
current mass and quark condensate, the logarithmic divergence of the 
condensate is precisely compensated by the vanishing of the bare 
current mass. The product is finite and coincides with the formula 
derived by Cahill and Gunner~\cite{cah96} and by Frank and 
Roberts~\cite{fra96}, implying that the GMOR relation (\ref{eq:1}) 
also holds in one-gluon-exchange models. 
\parindent=0pt

\bigskip 

The Euclidean quark condensate is defined by 
\be 
\langle \bar{q}q \rangle \; := \; - \lim _{x \rightarrow 0} 
\, \tr {\cal S} (x) \; , 
\label{eq:2} 
\en 
where ${\cal S} (x)$ is the quark propagator, which in general can be 
decomposed in momentum space as 
\be 
S(k) \; = \; \frac{i}{ Z (k^2) \kslash \, + \, i \Sigma (k^2) } 
\label{eq:3} 
\en 
At large momentum transfer $k$, the asymptotic behavior of the 
function $Z(k^2)$ and the self-energy $\Sigma (k^2)$ can be obtained
in perturbation theory augmented with operator-product corrections. 
$Z(k^2)$ rapidly approaches $1$, 
\be 
Z(k^2) \; \approx \; 1 \; + \; \; (\hbox{ logarithmic corrections}) \; , 
\label{eq:4a} 
\en
whereas $\Sigma (k^2)$ behaves like~\cite{pol76,mir93}, 
\be 
\Sigma (k^2) \approx \frac{ m_R (\mu  ) }{ \left[ 
\ln ( k ^2 / \mu ^2 ) \right] ^{d_m} } \, - \, 
\frac{ 4 \pi ^2 d_m }{ 3 k^2 } \frac{ \langle \bar{q} q \rangle _{OPE}
(\mu ) }{ \left[ \ln ( k ^2 / \mu ^2 ) 
\right] ^{1-d_m} } \; + \; \cdots \; , 
\label{eq:4} 
\en 
where $\mu $ is the renormalization scale, $m_R$ the renormalized 
current quark mass and $\langle \bar{q} q \rangle _{OPE} $ the 
quark condensate used as parameter in the OPE. 
$d_m$ is the anomalous mass dimension, which can be calculated in 
perturbative QCD, i.e. 
\be 
d_m \; = \; \frac{ 12 }{33 - 2 N_f} \; , 
\label{eq:5} 
\en 
with $N_f$ being the number of quark flavors. Inserting (\ref{eq:4}) 
into (\ref{eq:2}), one finds~\cite{rob94} in the chiral limit $m_R 
\equiv 0$ without resorting to a particular model 
\be 
\langle \bar{q}q \rangle \; := \; - 4 N_c \int _{k^2\leq \Lambda ^2} 
\frac{ d^{4}k }{ (2\pi )^4} \; 
\frac{ \Sigma (k^2) }{ Z^2 (k^2) k^2 + \Sigma ^2 (k^2) } \; = \; 
\langle \bar{q}q \rangle _{OPE} (\mu ) \, \left[ \ln 
\frac{ \Lambda ^2 }{ \mu ^2} \right] ^{d_m} \; + \; \ldots 
\label{eq:6} 
\en 
where $N_c$ is the number of colors and the dots indicate finite terms. 
This implies that the definition (\ref{eq:2}) in general produces 
a logarithmic divergent condensate (in the chiral limit\footnote{ 
The divergence of the condensate for $m_R \not= 0$ is even 
worse.}.) 

The crucial observation is that (\ref{eq:6}) does not rule out the 
definition of the quark condensate (\ref{eq:2}), since the 
GMOR relation requires the product of current mass and condensate. 
The generic behavior of the bare current mass is to vanish at large 
values of the UV-cutoff $\Lambda $ (see e.g.~\cite{co89}). 
One finds in QCD~\cite{yn83,mir93} 
\be 
m (\Lambda ) \; = \; \frac{ m_R (\mu ^2) }{ \left[ 
\ln \frac{ \Lambda ^2 }{ \mu ^2} \right] ^{d_m} } \; .
\label{eq:7} 
\en 
This directly leads to a cancelation of the logarithmic divergence 
in the product of current mass and condensate~\cite{mir93}, i.e. 
\be 
m ( \Lambda ) \, \langle \bar{q}q \rangle (\Lambda ) \; = \; 
m_R( \mu ) \, \langle \bar{q}q \rangle _{OPE} (\mu ) \; . 
\label{eq:8} 
\en 
This implies that the condensate appearing in the 
OPE is perfectly compatible with the one entering the GMOR relation. 

\bigskip 
In the rest of the letter, we will show that the mass formula, 
derived by Cahill and Gunner~\cite{cah96} and independently by 
Frank and Roberts~\cite{fra96}, coincides with the left hand side 
of (\ref{eq:8}). For this purpose, we resort to a particular model, 
where the self-energy is provided by the Dyson-Schwinger equation 
\be 
\Sigma (p^2) \; = \; m(\Lambda ) \, + \, \int _{k^2\leq \Lambda ^2} 
\frac{d^4k}{(2\pi )^4} \; 
D((p-k)^2) \, \frac{ \Sigma (k^2) }{ Z^2 (k^2) k^2 + \Sigma ^2 (k^2) } \; , 
\label{eq:9} 
\en 
where $\Lambda $ is the UV-regulator, which we will later take to 
infinity. The kernel 
\mbox{$D((p-k)^2)$} can be interpreted as Lorentz and color trace 
of the effective gluon-propagator. It is not necessary to specify 
$D((p-k)^2)$ 
for our argument here. We only demand that $D(k^2)$ matches the 
large momentum behavior known from perturbative QCD, i.e. 
\be
D(k^2) \; \approx \; \frac{ \hbox{ const. } }{ k^2 \, \ln k^2 / \mu ^2 } 
\; , \hbox to 2 cm {\hfil for \hfil } k^2 \gg \mu ^2 \; , 
\label{eq:9a} 
\en 
and that the function $D((p-k)^2)$ peaks at $p=k$. Here, we do not consider 
models which request a modification of the UV-renormalization in order 
to account for infra-red divergences.  
With the above standard 
assumptions, one observes that the ans\"atze (\ref{eq:4a}) and 
(\ref{eq:4}) for the asymptotic behavior of the self-energy and 
wave function are self-consistent solutions of the DS-equations. 
In particular, the momentum integration (\ref{eq:9}) 
is in fact UV-finite, and the cutoff dependence of the current 
mass in (\ref{eq:9}) is precisely that of (\ref{eq:7})~\cite{la96}. 

For later use, we also provide the self-energy 
$\Sigma _0(k^2)$ in the chiral limit, which satisfies 
\be 
\Sigma _0(p^2) \; = \; \int _{k^2\leq \Lambda ^2} 
\frac{d^4k}{(2\pi )^4} \; D((p-k)^2) \, 
\frac{ \Sigma _0(k^2) }{ Z^2 (k^2) k^2 + \Sigma _0^2 (k^2) } \; . 
\label{eq:10} 
\en 
For small deviations off the chiral limit ($m_R \ll \Sigma _0(0)$), one 
might neglect the change in the function $Z(k^2)$~\cite{rob94}  implying 
that the same function $Z(k^2)$ appears in (\ref{eq:9}) and 
(\ref{eq:10})\footnote{In the models excluded above (e.g.~\cite{sme91}) 
the wave function renormalization might be necessary to cover infra-red 
divergences. A more appropriate treatment of the function $Z(k^2)$ 
is requested in these models.}. 

The properties of the light pseudo-scalar mesons in the case of a small 
explicit breaking of chiral symmetry can be addressed by 
powerful techniques which step by step exploit the symmetry aspects 
and which became standard nowadays (for a review see e.g.~\cite{rob94}). 
Exploiting the chiral symmetry, it is possible to relate 
the quark self-energy to the Bethe-Salpeter vertex for the 
pion, i.e. 
\be 
P_0(k^2) \; = \; \frac{1}{f_\pi} \, \Sigma _0(k^2) \, \gamma _5 \; . 
\label{eq:11} 
\en 
The normalization 
is obtained by normalizing the charge of the pion with the help of 
the electro-magnetic form factor to unity. The full electro-magnetic 
vertex function at small momentum of the incoming photon, which enters 
the electro-magnetic form factor, and the full axial vector vertex function, 
needed to bring $f_\pi $ into the game, are unambiguously known 
from differential Ward identities. Expanding the pion's Bethe-Salpeter 
equation to leading order in $m_R$ (which is by assumption small to the 
scale set by $\Sigma _0(0)$) and $m_\pi ^2$ with the help of a 
derivative expansion, we find 
\be 
m_\pi ^2 \, f_\pi ^2 \; = \; 8 N_c \int _{k^2\leq \Lambda ^2} 
\frac{ d^{4}k }{ (2\pi )^4 } \; 
\Sigma _0^2(k^2) \, \frac{ \Sigma ^2 \, - \, \Sigma _0^2 }{ 
(Z^2 k^2 \, + \, \Sigma ^2) ( Z^2 k^2 \, + \, \Sigma _0^2) } \; . 
\label{eq:12} 
\en 
In this equality, the regulator can be safely removed ($\Lambda 
\rightarrow \infty $), since the integrand asymptotically decreases like 
$1/k^8$ (up to logarithmic corrections). 
Since the deviation of $\Sigma (k^2)$ off its chiral limit value 
$\Sigma _0(k^2)$ is proportional to the mass $m_R (\mu )$ which is the scale 
of explicit chiral symmetry breaking, we define 
\be 
\Sigma (k^2) \; = \; \Sigma _0(k^2) \, + \, m_R \, \sigma (k^2) \; . 
\label{eq:13} 
\en 
The asymptotic behavior of $\Sigma (k^2)$ and $\Sigma _0(k^2)$, 
deduced from (\ref{eq:4}), implies 
\be 
\sigma (k^2) \; \approx \; \frac{1}{ \left[ 
\ln ( k ^2 / \mu ^2 ) \right] ^{d_m} } \; , \hbox to 2cm 
{\hfil for \hfil } k^2 \gg \mu^2 \; . 
\en 
Inserting (\ref{eq:13}) into equation (\ref{eq:12}) yields to leading 
order in $m_R$ 
\be 
m_\pi ^2 \, f_\pi ^2 \; = \; 16 N_c \int \frac{ d^{4}k }{ (2\pi )^4 } \; 
\Sigma _0^3(k^2) \, \frac{ m_R \, \sigma (k^2) }{ 
( Z^2 k^2 \, + \, \Sigma _0^2 )^2 } \; . 
\label{eq:14} 
\en 
This result agrees with the findings of Cahill and Gunner~\cite{cah96} 
and of Frank and Roberts~\cite{fra96}. 
The main observation is that even for constant $\sigma (k^2)$ 
the momentum integration rapidly converges. 

In order to relate the right hand side of (\ref{eq:14}) to equation 
(\ref{eq:8}), we first expand the Dyson-Schwinger equation (\ref{eq:9}) 
to leading order in $m_R(\mu )$, i.e. 
\bea 
m_R \, \sigma (p^2) &=& m(\Lambda ) \; - \; \int _{k^2\leq \Lambda ^2} 
\frac{ d^{4}k }{ (2\pi )^4 } \; 
D((p-k)^2) \, \frac{ 2 \Sigma _0^2(k^2) \, m_R \, \sigma (k^2) }{ 
(Z^2 k^2 + \Sigma _0^2)^2 } 
\label{eq:15} \\ 
&+& \int _{k^2\leq \Lambda ^2} \frac{ d^{4}k }{ (2\pi )^4 } \; 
D((p-k)^2) \, \frac{ m_R \, \sigma (k^2) }{ Z^2 k^2 + \Sigma _0^2 } 
\; + \; {\cal O} (m_R^2) \; . 
\nonumber 
\ena 
Note that the series (\ref{eq:15}) with respect to powers of 
$m_R$ is well defined, since it can be verified that all appearing 
integrals are UV-finite. 

In order to relate the formula of Cahill and Gunner (\ref{eq:14}) to the 
standard GMOR relation, we multiply both sides of (\ref{eq:15}) by 
$$
\frac{ \Sigma _0 (p^2) }{ Z^2(p^2) p^2 + \Sigma ^2_0(p^2) } 
$$
and integrate over the momentum $p$, i.e. 
\be 
\int _{p^2\leq \Lambda ^2} 
\frac{ d^{4}p }{ (2\pi )^4 } \frac{ m_R \, \sigma (p^2) \Sigma _0 (p^2) }{ 
Z^2(p^2) p^2 + \Sigma ^2 _0(p^2) } \; = \; 
m(\Lambda ) \int _{p^2\leq \Lambda ^2} 
\frac{ d^{4}p }{ (2\pi )^4 } \; \frac{ \Sigma _0 (p^2) }{ 
Z^2(p^2) p^2 + \Sigma ^2 _0(p^2) } 
\label{eq:16} 
\en 
$$
- \; \int _{k^2\leq \Lambda ^2} 
\frac{ d^{4}k }{ (2\pi )^4 } \; \Sigma _0(k^2) \, 
\frac{ 2 \Sigma _0^2(k^2) \, m_R \, \sigma (k^2) }{ (Z^2 k^2 + 
\Sigma _0^2)^2 } 
\, + \, \int _{k^2\leq \Lambda ^2} 
\frac{ d^{4}k }{ (2\pi )^4 } \; \Sigma _0(k^2) \, 
\frac{ m_R \, \sigma (k^2) }{ Z^2 k^2 + \Sigma _0^2 } \; . 
$$
In the last line of (\ref{eq:16}), we have used the Dyson-Schwinger
equation (\ref{eq:10}) for the self-energy $\Sigma _0(k^2)$ in the 
chiral limit. The first and the last term of (\ref{eq:16}) cancel 
to yield 
\be 
m_R(\mu ) \int _{k^2\leq \Lambda ^2} \frac{ d^{4}k }{ (2\pi )^4 } \; 
\Sigma _0(k^2) \, 
\frac{ 2 \Sigma _0^2(k^2) \, \sigma (k^2) }{ (Z^2 k^2 + \Sigma _0^2)^2 } 
= m(\Lambda ) \int _{p^2\leq \Lambda ^2} 
\frac{ d^{4}p }{ (2\pi )^4 } \; 
\frac{ \Sigma _0 (p^2) }{ Z^2(p^2) p^2 + \Sigma ^2_0(p^2) } \; . 
\label{eq:17a} 
\en 
The divergence of the integral at the right hand side of (\ref{eq:17a}) 
in the limit $\Lambda \rightarrow \infty $ is precisely compensated 
by the vanishing bare current mass $m(\Lambda )$ to yield the 
finite result at the left hand side of (\ref{eq:17a}). 
We therefore obtain our main result 
\bea 
m_\pi ^2 \, f_\pi^2 &=& - 2 m_R(\mu ) \, \langle \bar{q} q \rangle _{OPE} 
\; = \; 2 \, \lim _{\Lambda \to \infty } m(\Lambda ) \; 
\lim _{x \to 0} \, \tr \, {\cal S} (x) 
\label{eq:18} \\ 
&=& 8 N_c m_R(\mu ) \int \frac{ d^{4}k }{ (2\pi )^4 } \; \Sigma _0(k^2) 
\, \frac{ 
2 \Sigma _0^2(k^2) \, \sigma (k^2) }{ (Z^2 k^2 + \Sigma _0^2)^2 } \; . 
\nonumber 
\ena
This result was numerically tested by Cahill~\cite{cah96p} using the model 
from~\cite{cah95} and by ourselves within the particular 
model~\cite{la96}. 

In conclusions, the cutoff-dependence of the bare current 
mass $m(\Lambda )$ precisely cancels the logarithmic divergence 
of the quark condensate which is defined via the trace of the 
quark propagator. The finite value of the product of current mass 
and quark condensate precisely yields the formula recently derived 
by Cahill and Gunner~\cite{cah96} and independently by Frank 
and Roberts~\cite{fra96}. We have therefore shown that there is 
no reason to conclude that the GMOR relation is wrong in 
the context of the one-gluon-exchange models. The above definition 
of the quark condensate 
is also compatible with its OPE definition via the 
asymptotic behavior of the self-energy. The GMOR relation is still 
satisfied.

\bigskip 
{\bf Acknowledgments: } 

We thank Reginald~T.~Cahill and Mannque Rho for encouragement and 
helpful comments, and R.~Alkofer for helpful remarks on the 
manuscript. We are also thankful to H.~Reinhardt for his continuous 
interest in this work as well as for support.

\end{document}